\documentclass[useAMS,usenatbib,fleqn]{mn2e}
\usepackage{amsmath, array, booktabs, xifthen, graphicx, color, xspace, url, amssymb, mathtools, enumitem, natbib, physics, listings, topcapt}
\usepackage[normalem]{ulem}

\usepackage{tikz,xcolor}
\usepackage{hyperref}
\hypersetup{colorlinks=true, linkcolor=black, filecolor=magenta, urlcolor=cyan, citecolor=black}

\definecolor{lime}{HTML}{A6CE39}
\DeclareRobustCommand{\orcidicon}{%
	\begin{tikzpicture}
	\draw[lime, fill=lime] (0,0) 
	circle [radius=0.16] 
	node[white] {{\fontfamily{qag}\selectfont \tiny ID}};
	\draw[white, fill=white] (-0.0625,0.095) 
	circle [radius=0.007];
	\end{tikzpicture}
	\hspace{-2mm}
}
\foreach \x in {A, ..., Z}{%
	\expandafter\xdef\csname orcid\x\endcsname{\noexpand\href{https://orcid.org/\csname orcidauthor\x\endcsname}{\noexpand\orcidicon}}
}
\def \aap{A\&A}

\def \apjl{ApJ}
\def \apjs{ApJS}

\newcommand{\reb}{{\sc \tt REBOUND}\xspace}
\newcommand{\rebx}{{\sc \tt REBOUNDx}\xspace}

\newcommand{\whckl}{{\sc \tt WHCKL}\xspace}

\newcommand{\bo}[1][]{%
    \ifthenelse{\equal{#1}{}}{\mathcal{O}}{\mathcal{O}\left(#1\right)}%
}

\NewDocumentCommand{\code}{v}{%
\texttt{{#1}}%
}

\newcommand{\fig}[1]{Fig.~\ref{#1}}
\newcommand{\eq}[1]{equation~(\ref{#1})}

\title[Stellar flybys and long-term stability]{On the long-term stability of the Solar System in the presence of weak perturbations from stellar flybys}
\date{Draft version: \today{}}

\author[Brown \& Rein]{Garett Brown$^{1,2}$\orcidA{} and 
    Hanno Rein$^{1,2,3}$\orcidB{}\\
$^1$ Department of Physical and Environmental Sciences, University of Toronto at Scarborough, Toronto, Ontario M1C 1A4, Canada,\\
$^2$ Department of Physics, University of Toronto, Toronto, Ontario, M5S 3H4, Canada,\\
$^3$ Department of Astronomy and Astrophysics, University of Toronto, Toronto, Ontario, M5S 3H4, Canada\\
}

\vspace{0.5\baselineskip}

\begin{document}
\maketitle

\begin{abstract}
The architecture and evolution of planetary systems are shaped in part by stellar flybys.
Within this context, we look at stellar encounters which are too weak to immediately destabilize a planetary system but are nevertheless strong enough to measurably perturb the system's dynamical state. 
We estimate the strength of such perturbations on secularly evolving systems using a simple analytic model and confirm those estimates with direct N-body simulations.
We then run long-term integrations and show that even small perturbations from stellar flybys can influence the stability of planetary systems over their lifetime.
We find that small perturbations to the outer planets' orbits are transferred between planets, increasing the likelihood that the inner planetary system will destabilize. 
Specifically, our results for the Solar System show that relative perturbations to Neptune's semi-major axis of order 0.1\% are strong enough to increase the probability of destabilizing the Solar System within 5 Gyrs by one order of magnitude.
\end{abstract}


%%%%%%%%%%%%%%%%%%%%%%%%%%%%%%%%%%%%%%%%%%%%%%%%%
%%%%%%%%%%%%%%%%%%%%%%%%%%%%%%%%%%%%%%%%%%%%%%%%%
\section{Introduction}
\label{sec:intro}
Since Newton's formulation of his universal law of gravitation, the long-term stability of the Solar System has been the subject of inquiry for centuries.
While a closed form analytic solution to the two-body problem exists, systems with more particles are not integrable and have no closed form analytic solution. 
Through perturbative expansions Laplace and Lagrange came up with secular equations with regular solutions \citep[for a comprehensive historical review see][]{2012arXiv1209.5996L}.
These early models were deterministic. 
However, using a computer assisted secular theory, \cite{Laskar1989} showed that the Solar System is chaotic.
This implies that any single solution to the set of N-body equations describing the Solar System only possesses specific predictive power for at most a few Lyapunov timescales ($\sim 100$~Myrs). 
Finally, using a large ensemble of direct N-body integrations, \cite{LaskarGastineau2009} showed that within a set of possible solutions to the Solar System's future evolution, there is a 1 per cent chance that Mercury's orbit will destabilize (leading to a collision or escape) within the next 5 Gyrs. 
Additional numerical and semi-analytic studies have mostly reconfirmed this result \citep{Zeebe2015, Mogavero2021, Abbot2021, Brown2020}.

The simulations mentioned above typically consider the Solar System in isolation, without any external perturbations. 
However, no system is completely isolated from the rest of the universe and in particular the effects of stellar flybys can be important depending on the stellar environment.
Different authors have considered such flybys under various circumstances: outer Solar System analogues \citep{Malmberg2011, Li2015, Stock2020}; 
planetary systems with one planet embedded in stellar clusters  \citep{Laughlin1998, Zheng2015}, systems with five planets with equal masses and separations \citep{Cai2017}, the Solar System excluding Mercury and Venus \citep{Dotti2019}; 
the late stages of the Solar System \citep{Zink2020}; and planetary systems with Kuiper belt and Oort cloud type objects \citep{PortegiesZwart2015}.
The general conclusion of these studies is that in the case of the Solar System in its current stellar neighbourhood, most flybys are weak and do not lead to an immediate dramatic destabilization of the system, and strong flybys are so rare that they are unlikely to occur within the Solar System's lifetime.

What is new in this paper is that we do not only look for immediate destabilization, but consider flybys which lead to weak perturbations and their implications for the very long-term evolution of the planetary system. 
Specifically, we derive a relationship between a stellar flyby and the effect it has on the probability that the system goes unstable within the next 5 Gyrs. 

To do this, we begin with a discussion of various stellar environments in Section~\ref{sec:flyby-effects}.
We then review the effect of a single stellar flyby and multiple stellar flybys on a single planet system's binding energy.
We discuss how perturbations to a system with two planets can be investigated through changes to the fundamental secular frequencies and how these frequencies are perturbed during flybys.
In Section~\ref{sec:numerics} we describe the numerical setup for our long-term ensemble of integrations of the Solar System.
We numerically verify the analytical estimate of the effects stellar flybys have on the secular frequencies of the Solar System and numerically determine the smallest perturbation which can affect the stability of the Solar System.
We close in Section~\ref{sec:conclusion} with the conclusions.

%%%%%%%%%%%%%%%%%%%%%%%%%%%%%%%%%%%%%%%%%%%%%%%%
%%%%%%%%%%%%%%%%%%%%%%%%%%%%%%%%%%%%%%%%%%%%%%%%
\section{Weak perturbations to planetary systems}
\label{sec:flyby-effects}

%%%%%%%%%%%%%%%%%%%%%%%%%%%%%%%%%%%%%%%%%%%%%%%%
\subsection{Stellar environments}
\label{sec:environments}

\begingroup
\setlength{\tabcolsep}{4pt}
\begin{table*}
   \centering
   \topcaption{A summary of the stellar densities $n$, typical velocities $\bar{v}$, and mass ranges $m_\star$ used to compute the data in \fig{fig:environments} and \fig{fig:successive}.
   Additionally, a summary of the expected time to a numerically resolvable flyby $\tau$, the expected time to a critical flyby $\tilde{\tau}_c$, and the expected lifetimes for the various stellar environments.
   }
   \begin{tabular}{@{}lccccccc} % Column formatting, @{} suppresses leading/trailing space
      \toprule
      \midrule
      Environment & Stellar Density & Typical Speed & Mass Range & Time to Flyby & Single Flyby & Multiple Flybys & Lifetime\\
      $\,$ & $n$ $[\mathrm{pc}^{-3}]$ & $\bar v$ [$\mathrm{km\ s}^{-1}$] & $m_\star$ $[M_\odot]$ & $\tau = \Gamma^{-1}$ [yrs] & $\tilde{\tau}_c$ [yrs] & $\tilde{\tau}_c$ [yrs] & [yrs]\\
      \midrule
      Local Neighbourhood & 0.14 & 26 & 0.01--10 & $9.76\cdot 10^{7}$ & $1.70\cdot 10^{11}$ & $(1.28\pm0.88)\cdot 10^{11}$ & - \\
      Open Cluster & 100 & 1 & 0.01--100 & $1.44\cdot 10^{8}$ & $1.22\cdot 10^{9}$ & $(5.11\pm4.61)\cdot 10^{8}$ & $\sim 10^{8}$ \\
      Globular Cluster & 1,000 & 10 & 0.01--1 & $1.72\cdot 10^{5}$ & $7.13\cdot 10^{7}$ & $(8.60\pm5.72)\cdot 10^{7}$ & - \\
      Milky Way Bulge & 50 & 120 & 0.01--10 & $5.49\cdot 10^{3}$ & $6.77\cdot 10^{8}$ & $(1.84\pm1.63)\cdot 10^{8}$ & - \\
      Milky Way Core & 10,000 & 170 & 0.01--10 & $1.17\cdot 10^{1}$ & $1.31\cdot 10^{6}$ & $(1.54\pm0.91)\cdot 10^{6}$ & - \\
      \bottomrule
   \end{tabular}
   \label{tab:summary}
\end{table*}
\endgroup

We want to draw statistical conclusions about the overall importance of stellar flybys for the Solar System. 
For that, we need to know the number density of stars, as well as the stellar mass and velocity distributions of the stellar environment. 
In this paper, we consider the following stellar environments.
The numerical parameters are also summarized in Table~\ref{tab:summary}.

\emph{Local Stellar Neighbourhood.}
We take the mass density to be $\hat n = 0.04\,M_\odot\,\text{pc}^{-3}$ \citep{Bovy2017} and combine it with the average mass from our initial mass function (IMF) for the neighbourhood, $\sim 0.3\,M_\odot$. 
We note that the typical velocities of local stars are comparable to Earth's orbital speed \citep{Bailer-Jones2018}.
% With a $\bar v \approx 26\,\mathrm{km\ s}^{-1}$, about $90 per cent$ of the samples from the Maxwell-Boltzmann distribution have $v_\star < 100\,\mathrm{km\ s}^{-1}$

\emph{Open Clusters.}
We consider an open cluster environment as it might be similar to the Solar System's stellar environment during its formation phase. 
Generally considered to remain gravitationally bound for about $100\,\mathrm{Myrs}$ \citep{Adams2010}, the variety, size, and density of open clusters can vary a lot, though \cite{PfalznerVincke2020} provide constraints on a possible cradle for the Sun.
They find that Solar System equivalents are more likely to be produced in areas where stellar densities range from $5\cdot 10^4\,\mathrm{pc}^{-3}< n_\mathrm{local} < 2\cdot 10^5\,\mathrm{pc}^{-3}$.
Since these densities are not necessarily typical of where stars spend the majority of their time in open clusters, \cite{Laughlin1998} suggest that for clusters that resemble isothermal spheres, the effective stellar density is a factor $\sim 10$ smaller than the maximum central density of the cluster.
Here, we adopt a commonly used stellar density of $n = 100\,\mathrm{pc}^{-3}$.
Relative velocities in typical open clusters are comparable to Neptune's orbital speed \citep{Adams2010, Malmberg2011}.

\emph{Globular Clusters.}
We consider the expected perturbation strengths that a Solar System analog would experience in a globular cluster, even though stars in such environments indicate they may be too metal poor to form such an analog~\citep{Fischer2005, Nascimbeni2012} and too dynamically hostile to maintain such an analog~\citep{Cai2019}.
Globular clusters are generally old \citep{Krauss2003} with small stars $\lesssim 0.85\,M_\odot$ \citep{Mann2019}, thus have the most restrictive mass range out of the environments we considered.
The velocity dispersion of globular clusters are similar to Jupiter's orbital speed \citep{Pryor1993, Mann2019}.
Additionally, the stellar density of globular clusters can vary by many orders of magnitude depending on the distance from the centre. We adopt a conservative static estimate representing a general distance $\sim 15\,\mathrm{pc}$ from the centre \citep{Jimenez-Torres2013}.

\emph{Milky Way Centre.}
Due to the extreme nature of the centre of the Milky Way, we follow the work of \cite{Jimenez-Torres2013} and divide the galactic centre into two regimes, the core (the inner $1\,\mathrm{pc}$) and the bulge (some $300\,\mathrm{pc}$ from the core).
The velocity dispersion in the galactic centre is much greater than any orbital velocity in the Solar System \citep{Jimenez-Torres2013, Valenti2018}.
As was shown by \cite{McTier2020}, with such high densities and velocities, as many as 4 in 5 stars in the galactic centre experience a flyby within $1000\,\mathrm{AU}$ within $1\,\mathrm{Gyrs}$.
We also show that flybys within this distance are common for systems in the galactic centre and show that they are also likely to disrupt planetary systems.

Incorporating the stellar density $n$, cross-section $\sigma$, and velocity dispersion $\bar v$, we can define the encounter rate $\Gamma$ as the rate in which stars enter a sphere of influence around the Sun \citep{Adams2010}:
\begin{equation}
    \Gamma =   n \sigma \bar v \,.
    \label{eq:gamma}
\end{equation}
We use this equation to calculate the average time it takes until a star comes within a certain distance from the Sun.
In the reference frame of the central star with mass $m_c$, a passing star with mass $m_\star$ begins with a relative velocity at infinity denoted by $v_\infty$.
If we consider the effects of gravitational focusing, the cross-section is \citep{BinneyTremaine2008, Malmberg2011, PortegiesZwart2015}
\begin{equation}
    \sigma = \pi b_\star^2\left(1 + \frac{2G (m_c + m_\star)}{b_\star v_\infty^2}\right)\,,
    \label{eq:focusing}
\end{equation}
where $b_\star$ is the impact parameter  of the incoming star and $G$ is Newton's gravitational constant. 
Note that for the vast majority of encounters, the impact parameter or the velocity are large and thus $\sigma \approx \pi b_\star^2$.

%%%%%%%%%%%%%%%%%%%%%%%%%%%%%%%%%%%%%%%%%%%%%%%%
\subsection{One planet}
\label{sec:one-planet}

We review the effects that a stellar flyby has on a binary. 
In our case the binary consists of a central star (the Sun) with mass $m_c$ and one planet with mass $m$, semi-major axis $a$, and eccentricity $e$.
For a passing star on a hyperbolic trajectory, the star's mass $m_\star$, eccentricity $e_\star$, impact parameter $b_\star$, relative velocity at infinity $v_\infty$, and perihelion $q_\star$ are related by \citep[e.g.][]{Spurzem2009}
\begin{equation*}
    e_\star = \sqrt{1 + \left(\frac{b_\star v_{\infty}^{2}}{G(m_c + m + m_\star)}\right)^{2}}
    \quad\textrm{and}\quad
    q_\star = b_\star \sqrt{\frac{e_\star - 1}{e_\star + 1}}\,.
    \label{eq:ecc}
\end{equation*}

Like other authors, we focus on adiabatic encounters where the velocity of the perturbing star at perihelion is less than that of the outermost planet.
We return to whether this is a good assumption below.
The binding energy of the central star and the planet is given by
\begin{equation}
    \varepsilon = -\frac{G m_c m}{2a}\,.
    \label{eq:e_simple}
\end{equation}
The change in binding energy of a binary system from an adiabatic encounter by a star on a hyperbolic trajectory is given by \citep{Roy2003, Heggie2006, Spurzem2009}
\begin{align}
    \Delta \varepsilon =&
    -\frac{\sqrt{\pi}}{8} G \mathcal{M}
    f_1(e_\star)
    \frac{a^2}{q_\star^3} K^{5/2}
    \exp\left[-\frac{2 K}{3}
    f_2(e_\star)\right] \notag \\
    & \cdot F(e, i_\star, \omega_\star, \Omega_\star, n t_0)\,,
\label{eq:de}
\end{align}
where
\begin{equation*}
    K = \sqrt{\frac{2 (m_c + m) q_\star^3}{(m_c + m_\star + m) a^3}}\quad\quad\textrm{and}\quad\quad 
     \mathcal{M} = \frac{m_c\,m_\star\,m}{m_c + m}\,.
\end{equation*}
The precise form of $F$ can be found in Appendix~\ref{sec:appendix}.
$F$ takes into account the orientation of the perturbing star's orbit relative to the elliptical orbit of the planet and is of order unity. 
The functions $f_1(e_\star)$ and $f_2(e_\star)$ which depend on the eccentricity of the flyby star are also given in Appendix~\ref{sec:appendix}.
In the parabolic case $f_1(1) = f_2(1) = 1$.

The relative change in binding energy due to a flyby can be used to calculate the relative change to the planet's semi-major axis, $\Delta a/a = -\Delta \varepsilon/\varepsilon$. 
This change in semi-major axis is what we refer to as the \emph{perturbation strength} of the stellar flyby.
We investigate the use of the angular momentum deficit (AMD) as an alternative metric in Appendix~\ref{sec:amd}.

Averaging over the incident angles, we obtain the average perturbation strength due to a flyby from \eq{eq:e_simple} and \eq{eq:de} so that
\begin{equation}
    \left|\frac{\Delta a}{a}\right| \simeq \left|\frac{\sqrt{\pi}}{2} \frac{m_\star}{m_c + m + m_\star} f_1(e_\star) K^{1/2} \exp\left[\frac{-2 K}{3} f_2(e_\star) \right]\right|.
    \label{eq:da/a}
\end{equation}
Note the exponential dependence on the perihelion of the flyby star (or more accurately the ratio of $q_\star/a$).
This implies that flybys will only have a significant effect if they come close, i.e. a distance comparable to the semi-major axis $a$.
From here we can probe the dependence of perturbation strength on the three main characteristics of perturbing stars: mass $m_\star$; distance $b_\star$; and speed $v_\infty$.
As an example, a flyby in the solar neighbourhood with $b_\star = 1000\,\mathrm{AU}$, $m_\star = 0.1\,M_\odot$, and $v_\infty = 40\,\mathrm{km\ s}^{-1}$ (so $q_\star \approx 999\,\mathrm{AU}$) will result in a relative change in the semi-major axis of Neptune of $|\Delta a/a| \approx 8.4\cdot 10^{-7}$

\begin{figure*}
    \centering
    \resizebox{0.99\textwidth}{!}{\includegraphics[trim=1.25cm 0.75cm 1.25cm 1.25cm]{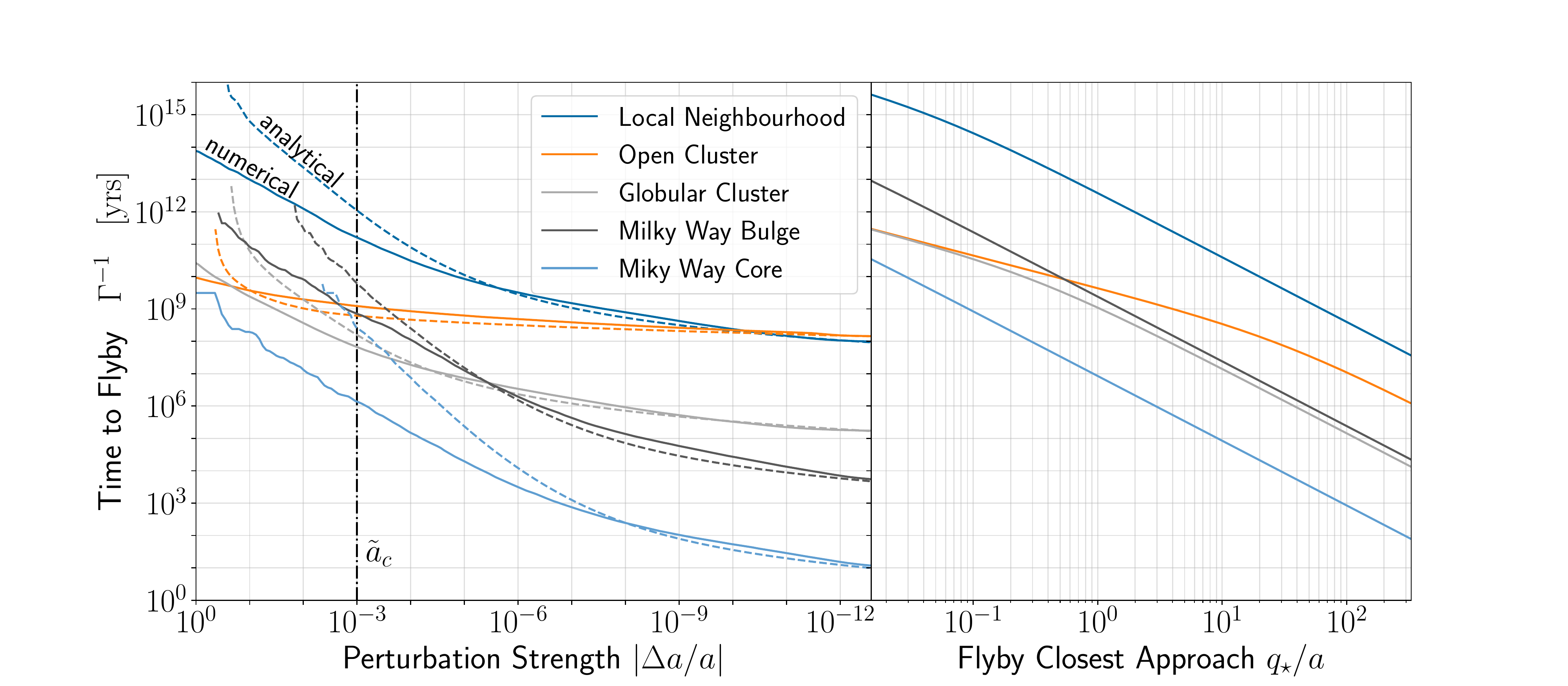}}
    \caption{The average time until a stellar flyby occurs for various stellar environments.
    On the left panel, we plot the time as a function of perturbation strength.
    On the right panel, we use the more traditional perihelion distance of the perturbing star normalized to Neptune's semi-major axis.
    The solid lines on the left panel are numerically measured changes to Neptune's semi-major axis, whereas the dashed lines are analytical estimates.
    The vertical dashed line at $\tilde{a}_c = 10^{-3}$ indicates where the perturbations are strong enough to affect the system's long term stability (see Section~\ref{sec:stability}).
    \label{fig:environments}
    }
\end{figure*}

We can obtain a more intuitive understanding of the importance of stellar flybys by calculating the average time it takes for an encounter with a specific perturbation strength to occur.
To do that, we sample the various stellar environments following the Monte Carlo approach of \cite{Zink2020} where for each flyby star we sample a mass, velocity, and impact parameter along with a random orientation centred on the Sun.
The initial mass function (IMF) we use to generate the stellar masses smoothly combines the IMF for single stars in equation~(17) by \cite{Chabrier2003} for stars less than one solar mass and the standard power-law IMF from \cite{Salpeter1955} for stars above one solar mass.
We sample stellar velocities from a Maxwell-Boltzmann distribution with a standard deviation $\bar v$ specific to the stellar environment (see Table~\ref{tab:summary}).
We sample impact parameters uniformly over the cross-sectional area. 
With $m_\star$, $v_\infty$, and $b_\star$, we compute the analytical estimate for the change in semi-major axis of Neptune orbiting the Sun using \eq{eq:da/a}.
We then take a subset of these samples and numerically measure the perturbation strength with 3-body simulations using \reb \citep{ReinLiu2012} to compare against the analytic estimate.
We compute a weighted histogram of the collected samples for both estimates, scaled according to the number of events expected in a given time.
We then cumulatively integrate the total number of flybys together such that a given flyby of perturbation strength $|\Delta a/a|$ or greater is expected after a given time (similarly for a flyby of certain perihelion or closer). 
The results are shown in \fig{fig:environments}.
    
The left panel of \fig{fig:environments} shows a comparison between analytical estimates and numerically measured perturbation strengths from stellar flybys for different environments.
Because we assume adiabatic encounters, we can see the greatest discrepancy between analytical and numerical results for impulsive flybys more often found in the environments with the largest velocity dispersion.
Note that the comparison between the analytical and numerical methods shows that for the majority of stellar environments the most frequent type of stellar flybys are in the adiabatic regime and result in very weak perturbations.
The flybys in open cluster environments in particular are almost always adiabatic.
The right panel of \fig{fig:environments} shows the perihelia from stellar flybys for different environments, which is the inverse of \eq{eq:gamma}.

Intuitively, the results in the left panel of \fig{fig:environments} can be understood by considering the three main environment characteristics: $n$, $m_\star$, and $\bar{v}$.
If we increase the density of the stellar region $n$ (keeping everything else constant) then the likelihood of a flyby increases, thus translating the curve in the figure down.
Similarly, if we increase the typical masses of the flyby stars $m_\star$, each individual perturbation would be more intense, translating the curve to the left.
Finally, if we change the velocity dispersion of the stellar environment $\bar{v}$ this changes the shape of the curve, with a shallow curve indicative of a very low velocity dispersion and a steep curve representative of a very high velocity dispersion.

To summarise, \fig{fig:environments} shows that the most common stellar flybys only cause weak perturbations to planetary systems.
Furthermore, we can see that the adiabatic assumption is more accurate for weaker encounters.
Finally, we can see that for secularly evolving systems, like the Solar System, critical perturbations $|\Delta a/a| > 10^{-3}$ (see Section~\ref{sec:stability}) are rare compared to the lifetime of the system.
The numerical values for the rarity of these critical perturbations is also shown in the Single Flyby column of Table~\ref{tab:summary}.

%%%%%%%%%%%%%%%%%%%%%%%%%%%%%%%%%%%%%%%%%%%%%%%%
\subsection{Successive flybys}
\label{sec:successive-flybys}

\begin{figure*}
    \centering
    \resizebox{0.8\textwidth}{!}{\includegraphics[trim=1.75cm 0.75cm 0.75cm 0cm]{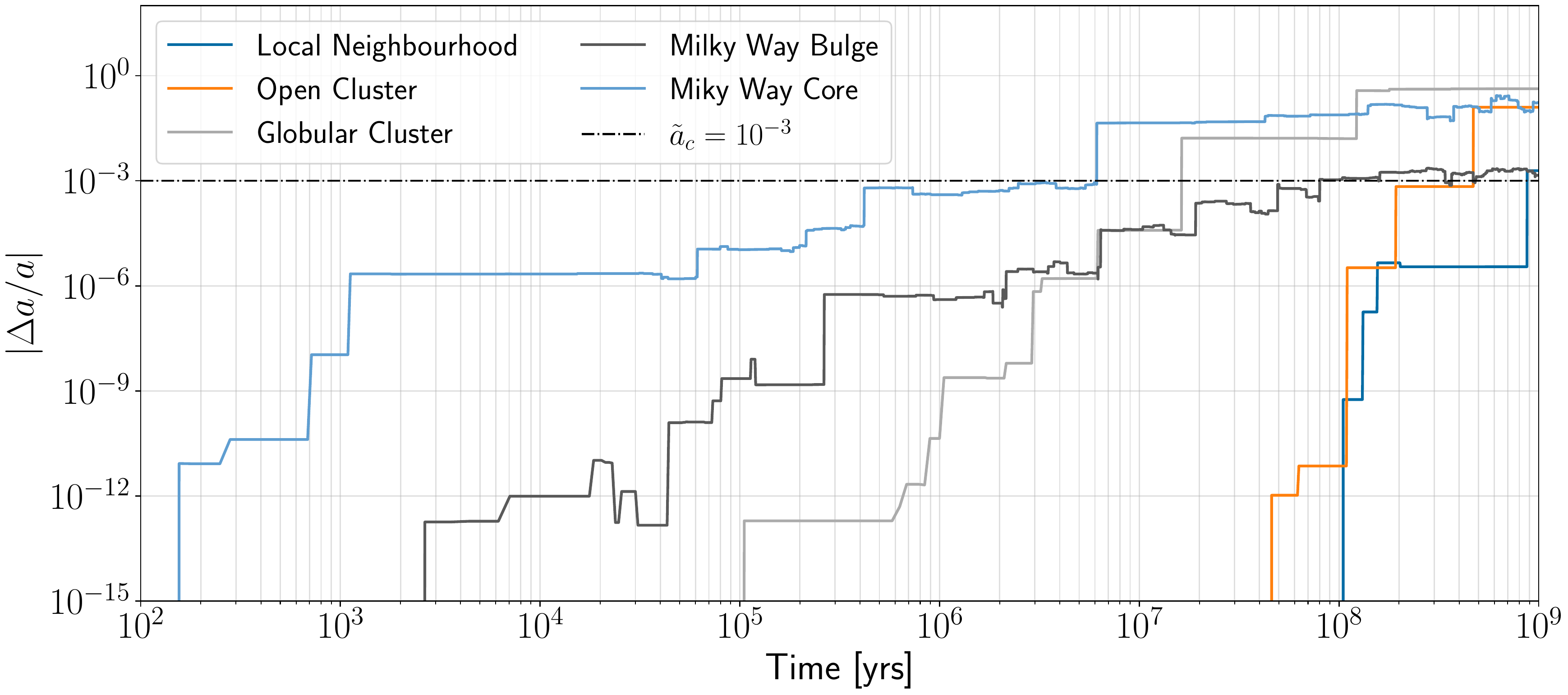}}
    \caption{The impact of successive flybys on the binding energy (semi-major axis) of a Sun-Neptune system.
    The horizontal dashed line at $\tilde{a}_c = 10^{-3}$ indicates where the perturbations are strong enough to measurably affect the system's long term stability (see Section~\ref{sec:stability}). The most important feature is that the largest relative change is the most relevant. We also observe that the perturbation to the systems generally increase overtime and the effects are akin to L\'evy flight.
    \label{fig:successive}
    }
\end{figure*}

We next consider the effects of stellar flybys on star-planet binary systems by simulating successive flybys to the same system.
The importance of successive flybys on the resulting architecture of planetary systems has recently been shown in multiple studies \citep{Cai2017, Stock2020}.
We begin with the same setup and Monte Carlo approach as Section~\ref{sec:one-planet} for a star-planet binary system where a Neptune-analogous planet orbits around a solar mass star.
We simulate the evolution of one planet orbiting a star for 1~Gyrs. 
We divide the time into 1~Myr pieces and determine the number of flybys expected to occur for a given environment in each million years using \eq{eq:gamma}. 
For each flyby that occurs, we randomly draw the three main characteristics of the perturbing stars (mass, velocity, and impact parameter) and simulate the flyby the same way we did in Section~\ref{sec:one-planet}.
We then assume that the resulting orbital parameters of the planet will remain constant until the next flyby and we repeat the processes of randomly drawing and simulating flybys for the duration of the 1~Gyrs.

We simulate 1024 star-planet systems in each stellar environment and calculate the average time until  $|\Delta a/a| > 10^{-3}$ has been reached. 
The results are in Table~\ref{tab:summary} under the Multiple Flybys column.
When assuming that the star-planet system remains in an environment with a constant stellar density, the cumulative effects of successive flybys do not significantly change the expected time for a system to experience a critical level perturbation.
A comparison between the columns in Table~\ref{tab:summary} for a single flyby and multiple flybys shows that on average, successive flybys increase the likelihood of a critical perturbation by at most about a factor of four (again, assuming a constant stellar density).
This is because the effects of successive flybys are dominated by the most significant flybys.
Note that this is consistent with the results shown in Fig.~\ref{fig:environments} which show that the kicks planetary systems experience are not all equal, but follow a distribution that includes rare but large kicks, reminiscent of a L\'evy flight.

\fig{fig:successive} shows one sample out of each set of 1024 simulations illustrating the typical evolution of the relative change in semi-major axis (compared to the original state) of a Neptune-like planet for each of the five stellar environments discussed above.
The general trend for systems undergoing successive flybys is to deviate more and more from their initial state, but not in a purely random walk way where the effects grow as $\sqrt{t}$.
The evolutionary paths for these systems are suggestive of a L\'evy flight, with long periods of little change, interrupted by sudden jumps when a strong encounter occurs (note the log-log scale). 
For dense environments, the systems are more quickly pushed from their initial states compared to less dense environments such as the Local Neighbourhood.
Here we acknowledge some of the limitations of our approach for simulating successive flybys. 
Since we assume a constant stellar density throughout the entire 1~Gyrs evolution, the particular path that a system might follow through a stellar environment will likely pass through more and less dense regions.
This approach of adopting typical and conservative values for the parameters of each stellar environment does not take into account the continuity of stellar density. 
While this may change the overall likelihood by more than a factor of four, the qualitative features seen \fig{fig:successive} remain the same.
Mainly, the changes to the systems are dominated by the most significant flybys.

%%%%%%%%%%%%%%%%%%%%%%%%%%%%%%%%%%%%%%%%%%%%%%%%
\subsection{Two planets}
\label{sec:linapprox}

Long-term variations of planetary orbits in the Solar System are caused by the cumulative effects of planet-planet interactions, general relativity, tides, and other external perturbations.
Because there are no significant mean motion resonances between the planets of the Solar System, the dynamics (especially the inner planets) are dominated by secular dynamics, i.e. the shape of the orbits including semi-major axes, eccentricities, inclinations, perihelia, and ascending nodes, but not by the planets' phases.
The frequencies of oscillations in perihelia (the rotation of the orbit in its orbital plane) and the ascending node (the rotation of the plane of the orbit in space) are referred to as secular frequencies.
The dominant mechanism for instability in the Solar System emerges from the overlap of secular resonances (see Section~\ref{sec:secular-resonances}).
For this reason, we will discuss how secular frequencies change if the Solar System is perturbed by a flyby.

For a two planet system, with a semi-major axis ratio of the inner and outer planets $\alpha = a_{1}/a_{2}$, there are two secular modes associated with perihelion precession having eigenfrequencies \citep{Murray1999},
\begin{equation*}
g_1 \simeq \frac{3}{4}\mu_2 n_1 \alpha^{3}
\quad\textrm{and}\quad
g_2 \simeq \frac{3}{4}\mu_{1}n_{2}\alpha^{2}
\end{equation*}
assuming $\alpha\ll 1$, with reduced masses $\mu_{1} = m_{1}/(m_{c} + m_{2})$, $\mu_{2} = m_{2}/(m_{c} + m_{1})$, and mean motions $n_{j}^2 = G(m_c + m_j)/a_{j}^{3}$.
With this dependence of the eigenfrequencies on the semi-major axes, we can determine the relative change in secular frequencies based on the change in semi-major axes of the planets. 
Namely,
\begin{equation*}
\left|\frac{\Delta g_1}{g_1}\right| = \left|\frac{3}{2}\frac{\Delta a_1}{a_1} - 3\frac{\Delta a_2}{a_2}\right|
\;\;\textrm{and}\;\;
\left|\frac{\Delta g_2}{g_2}\right| = \left|2\frac{\Delta a_1}{a_1} - \frac{7}{2}\frac{\Delta a_2}{a_2}\right|.
\end{equation*}
Thus, we find that to first order $\Delta g/g \propto \Delta a/a$. 

In the event of a flyby, the exponential dependence of $a_j$ in \eq{eq:da/a} implies that $\left|\Delta a_j/a_j\right|$ will be smaller for the inner planet than for the outer planet.
Therefore, the change in secular modes will be dominated by the changes to the semi-major axis of the outermost planet.
Additionally, numerical studies involving stellar flybys have shown that changes to the architecture of a planetary system correlates with the number of planets and the initial semi-major axis of the outermost planet \citep{Stock2022}.
Note that we now have a straightforward scaling relationship that goes from encounter parameters ($m_\star$, $b_\star$, $v_\infty$) to $|\Delta g_j/g_j|$.
As an example, consider a flyby in the solar neighbourhood with $b_\star = 1000\,\mathrm{AU}$, $m_\star = 0.1\,M_\odot$, and $v_\infty = 40\,\mathrm{km\ s}^{-1}$.
If we consider only Jupiter and Neptune, the relative change in secular frequencies is $|\Delta g/g| \approx 2.5\cdot 10^{-6}$.

%%%%%%%%%%%%%%%%%%%%%%%%%%%%%%%%%%%%%%%%%%%%%%%%
\subsection{More than two planets}
\label{sec:secular}

Above we have estimated the effects that a flyby has on the secular frequencies of a two planet system.
We now consider the relative change in secular frequencies in the case of more than two planets.
We will show with numerical simulations that stellar flybys cause relative changes to all secular modes on the same order of magnitude as the relative changes to Neptune's semi-major axis, i.e. $\Delta g_j/g_j \propto \Delta a_8/a_8$.
Note that secular modes can never be exclusively associated with one single planet.
So although one might refer to $g_1$ as Mercury's secular mode, it is really a mode of the entire Solar System, and all modes are coupled to all planets.

\begin{figure}
    \centering
    \resizebox{0.99\columnwidth}{!}{\includegraphics[trim=1cm 1cm 1cm 1cm]{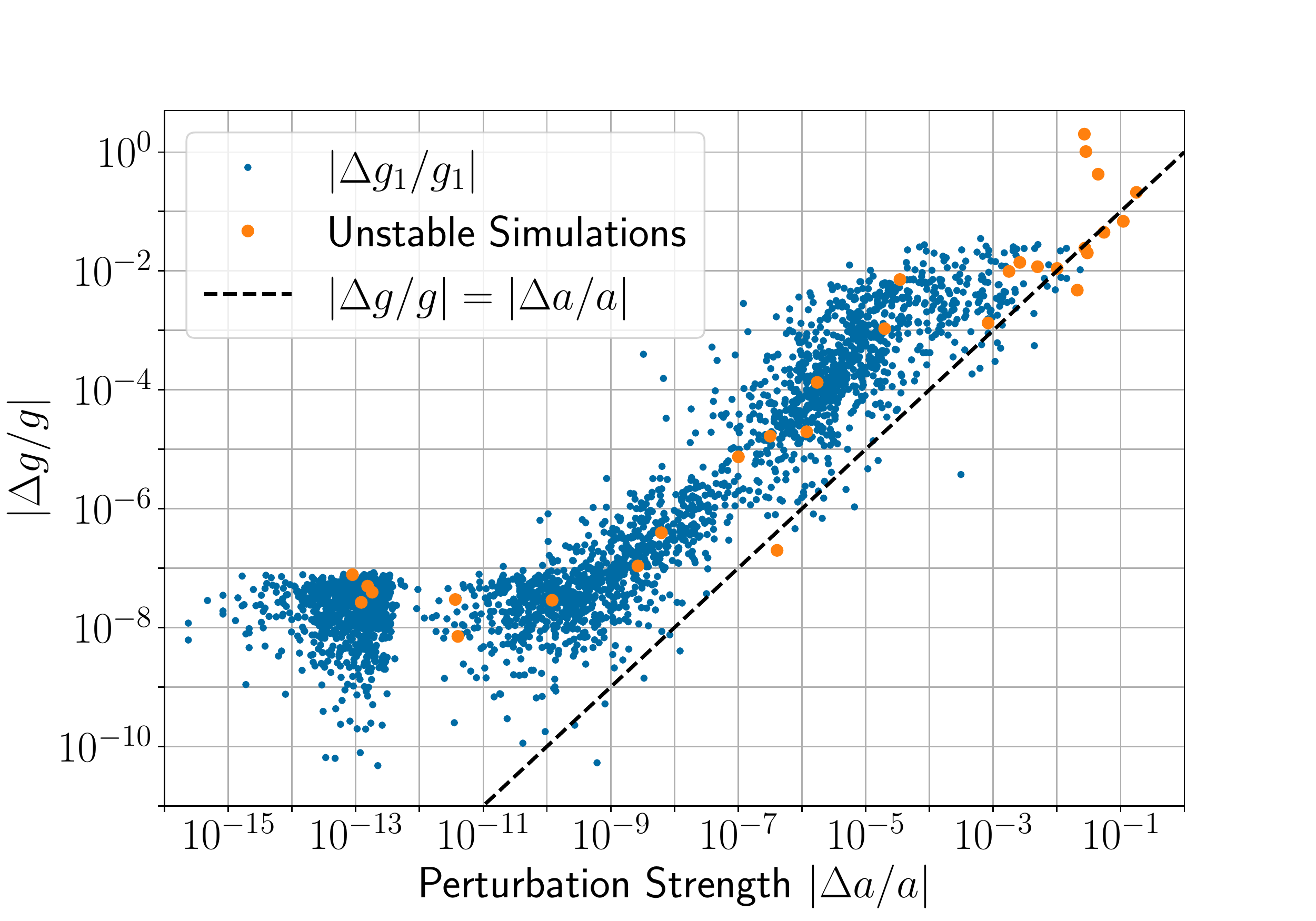}}
    \caption{Changes to the secular mode $g_1$ as a function of the perturbation strength $\Delta a/a$, measured in Neptune's orbit.
    Perturbations with $\Delta a/a<10^{-11}$ cannot be resolved due to finite floating point precision.
    \label{fig:dg-vs-da}
    }
\end{figure}

To calculate the secular frequencies numerically, we follow the procedure developed by \cite{Laskar1988, Laskar1990, Laskar1993, Laskar2003} and known as modified frequency analysis. See also the work by \cite{Sidlichovsky1997}.
We integrate an ensemble of 2,880 stellar flybys to the Solar System (for the precise numerical setup see Section \ref{sec:numerics}).
We measure the change in secular frequencies over 20~Myrs compared to an unperturbed case.
The results are shown in \fig{fig:dg-vs-da}.

One can see that the relationship between the relative change in $g_1$ and the numerically measured relative change in the semi-major axis of Neptune from stellar flybys (what we call the perturbation strength) is linear except where we reach the numerical floor at $\Delta a_8/a_8<10^{-11}$, and where the perturbations are of order unity.
The other modes, $g_2, \ldots, g_8$ similarly show this linear behaviour.
Thus, in spite of the rough estimates we made in Section~\ref{sec:flyby-effects}, \fig{fig:dg-vs-da} shows that even though a stellar flyby is very unlikely to directly alter the orbit of Mercury in any significant way, secular interactions will eventually propagate perturbations of the outer planets' orbits to Mercury's orbit.
The timescale for these changes to propagate from the outer planets to the inner planets is the secular timescale. 
The secular timescales for the Solar System are in the hundreds of thousands to tens of millions of years \citep{LithwickWu2011}.
We note that changes in the secular frequencies can be calculated with high accuracy but this does not provide any immediate new insight into the physical state of the system for changes smaller than $\sim 10^{-4}$ \citep{ReinBrownTamayo2019}.

%%%%%%%%%%%%%%%%%%%%%%%%%%%%%%%%%%%%%%%%%%%%%%%%
%%%%%%%%%%%%%%%%%%%%%%%%%%%%%%%%%%%%%%%%%%%%%%%%
%%%%%%%%%%%%%%%%%%%%%%%%%%%%%%%%%%%%%%%%%%%%%%%%
\section{Long-term stability}
\label{sec:numerics}

%%%%%%%%%%%%%%%%%%%%%%%%%%%%%%%%%%%%%%%%%%%%%%%%%
\subsection{Numerical methods}
\label{sec:setup}

For all our simulations, we use NASA JPL Horizons data at the J2000 epoch as initial conditions for the Solar System (Sun and $8$ planets).
We integrate simulations forward in time using \reb \citep{ReinLiu2012} and the Wisdom-Holman integrator \cite{WisdomHolman1992} with symplectic correctors and the lazy implementation of the kernel method, \whckl \citep{Wisdom1996}.
This integrator is well suited to provide highly accurate results for secularly evolving systems \citep{ReinTamayoBrown2019}. 
We include general relativistic corrections with \rebx \citep{Tamayo2020} using the \texttt{gr\_potential} module.
We used a fixed timestep of $dt = 8.062\,\mathrm{days}$.

%%%%%%%%%%%%%%%%%%%%%%%%%%%%%%%%%%%%%%%%%%%%%%%%

%%%%%%%%%%%%%%%%%%%%%%%%%%%%%%%%%%%%%%%%%%%%%%%%%
\subsection{Solar System stability}
\label{sec:stability}

In this section we present the stability results of an ensemble of 2,880 long-term integrations of the Solar System. 
Each simulation begins with a flyby star passing along on a random hyperbolic trajectory.
The flybys are sampled from an open cluster environment and we record the perturbation strength for each of them.
We then integrate the simulations for $4.8$ Gyrs, or until a collision or escape event occurs.

We divide the simulations into a control group, where the perturbations are too small to be resolved from numerical noise, and an experimental group, where relative perturbations range from numerical noise to near unity.
From the control group, we find that 4 of the 960 simulations (0.42 per cent) ended with a Mercury-Venus collision because of a dramatic increase in the eccentricity of Mercury. 
Within the experimental group, we find that 26 of the 1,920 simulations (1.35 per cent) ended in instability --- 20 were collisions between Mercury and Venus, one was a collision between Earth and Mars, two resulted in the escape of Uranus, two ended in the escape of Neptune, and one finished with the escape of Mercury. A more detailed breakdown of the results can be found in Appendix~\ref{sec:details}.

Let us consider more closely the effect of the different strengths of stellar flybys on the Solar System.
\begin{figure}
    \centering
    \resizebox{0.99\columnwidth}{!}{\includegraphics[trim=1cm 1cm 1cm 1cm]{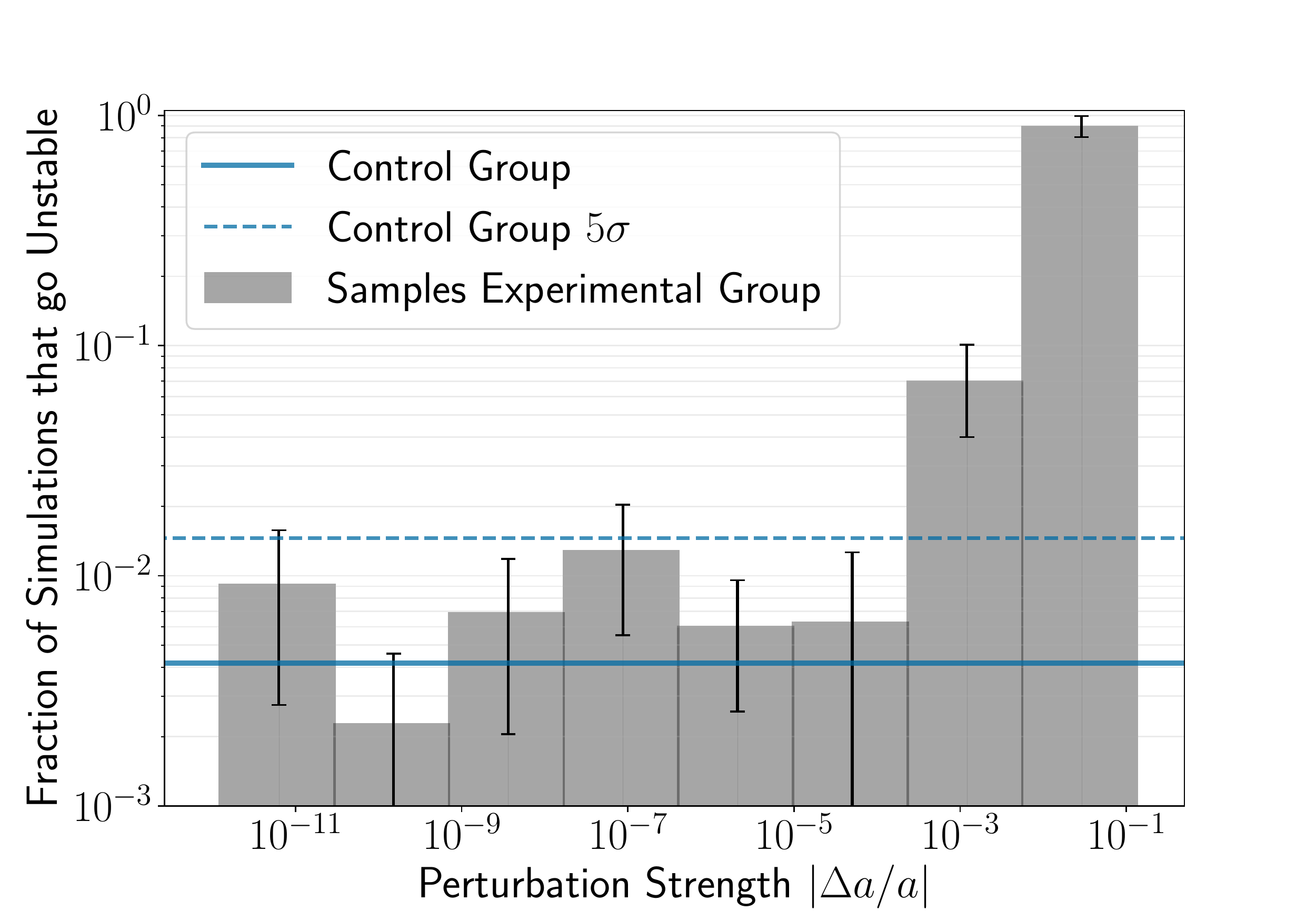}}
    \caption{The fraction of simulations that lead to an instability before 5 Gyrs grouped by perturbation strength $\Delta a/a$. 
     For perturbation strengths larger than $\tilde{a}_c = 10^{-3}$ the fraction is significantly higher (more than $5\sigma$) than in the control group.
    \label{fig:stability-by-strength}
    }
\end{figure}
\fig{fig:stability-by-strength} shows the fraction of systems that go unstable as a function of the numerically measured perturbation strength, $\Delta a/a$.
There is a clear qualitative change in the fraction of unstable simulations starting at around $|\Delta a/a| \simeq 10^{-3}$ when the fraction of unstable simulations is more than five standard deviations beyond the mean of the control group. 
We thus define the critical perturbation strength to be $\tilde{a}_c = 10^{-3}$.
Perturbations stronger than $\tilde{a}_c$ can significantly alter the dynamical state of the Solar System in a way that affects its long term stability.
On the other hand, the Solar System appears to be robust to perturbations of less than $\tilde{a}_c$ as they do not significantly alter the probability of an instability occurring.

Note that nearly 100 per cent of the simulations go unstable around $|\Delta a/a| \simeq 10^{-1}$, corresponding to Neptune's semi-major axis changing by about $3\,\mathrm{AU}$.
Our analytic approximations break down in that regime as the perturbations are no longer weak.
An example of a flyby in the solar neighbourhood that results in this strong of a perturbation would be $m_\star = 1\,M_\odot$, $b_\star = 250\,\mathrm{AU}$, and $v_\star = 20\,\mathrm{km\ s}^{-1}$ (so $q_\star \approx 245\,\mathrm{AU}$).

As we did not carry out any simulations beyond a collision or escape, the remaining systems could be stable for many Gyrs, or could entirely destabilize \citep{LaskarGastineau2009}.

%%%%%%%%%%%%%%%%%%%%%%%%%%%%%%%%%%%%%%%%%%%%%%%%
\subsection{Secular resonances}
\label{sec:secular-resonances}
\begin{figure}
    \centering
    \resizebox{0.99\columnwidth}{!}{\includegraphics[trim=1.25cm 1.25cm 1.25cm 1.25cm]{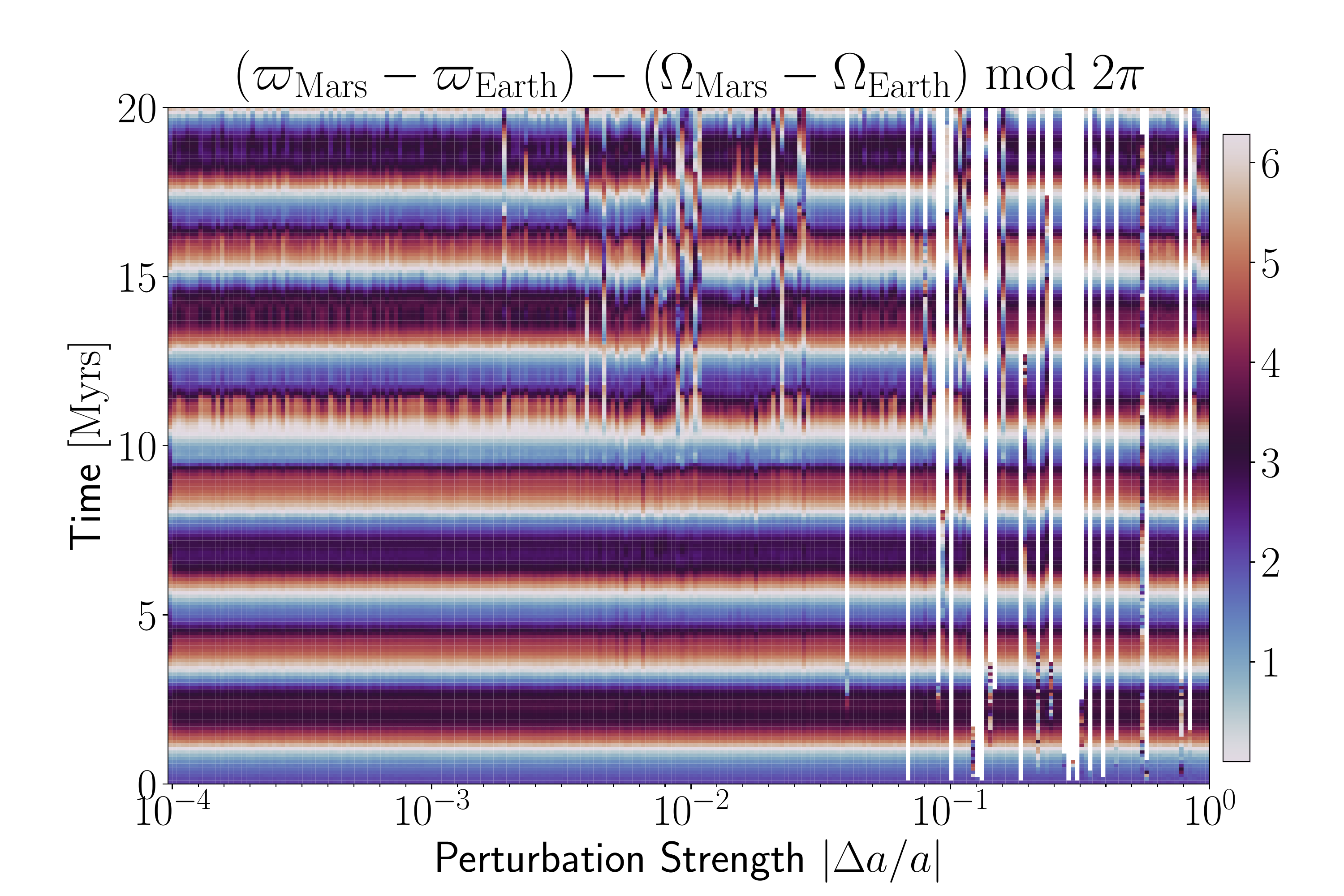}}
    \caption{Secular resonant angle $[1,-1]$ between Earth and Mars \citep{LithwickWu2011} as a function of perturbation strength and time for 240 different simulations. 
    Even on this timescale, small perturbation to Neptune's orbit of $\Delta a/a\gtrsim 10^{-3}$ can alter the resonant structure of the inner Solar System.
    Almost instantaneous instability events (vertical white lines) can occur at $\Delta a/a\gtrsim 10$ per cent.
    \label{fig:earth-mars}
    }
\end{figure}

As we've argued in Section~\ref{sec:secular}, perturbations to the orbit of the outermost planet have effects on all secular frequencies.
This is important because it can push planets in or out of secular resonances with each other \citep{Laskar2000, Zakamska2004}.
For example, a resonance involving the longitude of perihelion and the longitude of the ascending node of Earth and Mars has been identified as a possibly important secular resonance in the inner Solar System \citep{Laskar1992, SussmanWisdom1992, LithwickWu2011}.
This resonance is related to the secular frequencies $g_3$ and $g_4$.
These frequencies are already close to a commensurability in the present-day unperturbed Solar System with a relative difference of only $\sim 3.1\cdot 10^{-2}$  \citep{Laskar2011, ReinBrownTamayo2019}.
In our long-term N-body simulations, we see correlations of this secular resonances with eccentricity pumping and, eventually, planetary collisions.
Another resonance, and in fact the one most correlated with instability in our simulations, is the $g_1 - g_5$ secular resonance between Mercury and Jupiter \citep{Mogavero2021}.
When the Solar System falls into this resonance, the eccentricity of Mercury increases on short timescales \citep{Batygin2015}.

To investigate this effect further we ran an additional 240~N-body simulations.
We initialize these simulations by first integrating the present day Solar System a random time (between 0 and 10,000~years) forward in time to randomize the orbital phases.
Rather than perturbing the system with a flyby we artificially perturb the orbital elements of Neptune by a small amount.
Each simulation is then integrated for 20 Myrs, or until a collision or escape event occurs.

In \fig{fig:earth-mars}, we plot the resonant angle $[1, -1]$ which is related to the $g_3$ and $g_4$ secular frequencies \citep[see][for details]{LithwickWu2011} as a function of time and perturbation strength.
We only show one angle for this illustration, but we note that other angles show very similar behaviour. 
In the unperturbed state (far left of the plot), the system is not in resonance in the first 20~Myrs as the angle is circulating.
However, for larger perturbation strengths, the resonant angle can be pushed into a librating state, at least temporarily.
This can be seen in the figure as the colour remains the same for an extended period of time compared to some of the neighbouring simulations which carry on normally. 
\fig{fig:earth-mars} shows that clear deviations from the normal secular structure of the Solar System begin with relative changes around $\tilde{a}_c$ and become catastrophic around $|\Delta a/a| \simeq 10^{-1}$, consistent with the results from our long term simulations.
Furthermore, one can see that about half of the simulations in the range $10^{-1} \leq |\Delta a/a| \leq 10^{0}$ remain stable for the first 20~Myrs following the perturbation. 
Note that we are now in a highly non-linear regime.
The perturbations change the secular dynamics of the Solar System significantly and we expect many of the simulations which are stable on short timescales will destabilize after a few secular timescales as energy and angular momentum are exchanged through planet-planet interactions.

%%%%%%%%%%%%%%%%%%%%%%%%%%%%%%%%%%%%%%%%%%%%%%%%%
%%%%%%%%%%%%%%%%%%%%%%%%%%%%%%%%%%%%%%%%%%%%%%%%%
%%%%%%%%%%%%%%%%%%%%%%%%%%%%%%%%%%%%%%%%%%%%%%%%%

\section{Discussion and conclusions}
\label{sec:conclusion}

In this paper, we explored the sensitivity of Solar System stability due to weak perturbations such as those resulting from distant stellar flybys.
Our numerical long term integrations show that the Solar System's stability is not affected by perturbations as long as they change the planets' orbital parameters by less than $0.1$ per cent ($\tilde{a}_c = 10^{-3}$).
We showed that for stellar environments similar to our local neighbourhood, flybys with such perturbation strengths only occur once every $100$~Gyrs.
However, for stars which are part of more dense environments, strong encounters are much more likely.
We estimate that encounters with $\tilde{a}_c > 10^{-3}$ will occur during a system's lifetime and likely play an important role in shaping such planetary systems.

We also explored the evolution of star-planet systems within various stellar contexts as they undergo multiple successive flybys.
The effects of successive flybys behave like a L\'evy flight where the outcome is mostly set by the largest perturbation, rather than the cumulative effect of equal perturbations in a random walk. 
Thus, in most cases it is sufficient to consider only the strongest perturbation to see whether the stability of a system is affected.

We derived a set of equations which allow us to estimate the effects from a stellar flyby on the semi-major axes of planets and the secular frequencies of planetary systems. 
These estimates are working well for weak (or adiabatic) encounters, but break down for strong encounters.
We've shown that all secular modes are affected during flybys, i.e. not only those typically associated with the outer planets.

The Solar System appears to be in a somewhat stable region of parameter space.
The path to instability is ultimately determined by stochastic processes and chaos \citep{Mogavero2022}. 
We argue that the critical perturbation strength to affect the long-term stability of the Solar System is $\tilde{a}_c = 10^{-3}$.
Perturbations stronger than $\tilde{a}_c$ lead to changes in the secular frequencies of the order of $\Delta g/g \approx 10^{-3}$.
Such perturbations are strong enough to affect the secular dynamics and push planets in and out of resonances which ultimately changes the probability of the system going unstable.
We note there is of course no one-to-one correspondence between changes to the secular frequencies and dynamical instability because of the chaotic nature of the system.

We also note that our study did not include stellar binaries, even though as many as half of stars are part of a binary system \citep{Chabrier2003} and more work needs to be done to better estimate the occurrence rate of such encounters along with their impact on perturbation strength. We anticipate it would increase the strength of the perturbations by about 2 to 3.6 times \citep{Li2015} and thus decrease the occurrence of a critical flyby in the Local Neighbourhood to about once every 64 to 75~Gyrs.
We also did not investigate the effects that Planet 9 \citep{Batygin2019} would have on the Solar System's stability in the context of a stellar flyby. 
Given its proposed distance from the Sun, we expect it to experience the largest relative change from a stellar flyby.
However, we also expect the secular coupling between Planet 9 and the rest of the Solar System to be weak.

Whereas our analytical estimates are general and apply to any planetary system, our numerical simulations focused on the Solar System, the system for which we have the most accurate ephemeris. 
A natural direction for future work would be to look more closely at the effects of weak stellar flybys on exoplanet systems.
 
%%%%%%%%%%%%%%%%%%%%%%%%%%%%%%%%%%%%%%%%%%%%
%%%%%%%%%%%%%%%%%%%%%%%%%%%%%%%%%%%%%%%%%%%%
\section*{Data availability}
The data underlying this article will be shared on reasonable request to the corresponding author.

%%%%%%%%%%%%%%%%%%%%%%%%%%%%%%%%%%%%%%%%%%%%
%%%%%%%%%%%%%%%%%%%%%%%%%%%%%%%%%%%%%%%%%%%%
\section*{ORCID iDs}
Garett Brown \orcidA{} \href{https://orcid.org/0000-0002-9354-3551}{https://orcid.org/0000-0002-9354-3551}
Hanno Rein \orcidB{} \href{https://orcid.org/0000-0003-1927-731X}{https://orcid.org/0000-0003-1927-731X}
 
%%%%%%%%%%%%%%%%%%%%%%%%%%%%%%%%%%%%%%%%%%%%
%%%%%%%%%%%%%%%%%%%%%%%%%%%%%%%%%%%%%%%%%%%%
\section*{Acknowledgments}
We are very grateful to Maxwell Xu Cai for a helpful review which significantly improved the substance and quality of this paper.
We would like to thank Scott Tremaine, Sam Hadden, Dan Tamayo, and Nathan Sandholtz for useful discussions. 
This research has been supported by the NSERC Discovery Grants RGPIN-2014-04553 and RGPIN-2020-04513.
This research was made possible by the open-source projects 
\texttt{Jupyter} \citep{jupyter}, \texttt{iPython} \citep{ipython}, \texttt{matplotlib} \citep{matplotlib, matplotlib2}, and \texttt{GNU Parallel} \citep{gnuparallel}.
This research was enabled in part by support provided by Compute Canada (www.computecanada.ca).
Computations were performed on the Niagara supercomputer \citep{SciNet2010, Ponce2019} at the SciNet HPC Consortium (www.scinethpc.ca). 
SciNet is funded by: the Canada Foundation for Innovation; the Government of Ontario; Ontario Research Fund - Research Excellence; and the University of Toronto.

%%%%%%%%%%%%%%%%%%%%%%%%%%%%%%%%%%%%%%%%%%%%%%%
%%%%%%%%%%%%%%%% Bibliography  %%%%%%%%%%%%%%%%
%%%%%%%%%%%%%%%%%%%%%%%%%%%%%%%%%%%%%%%%%%%%%%%

\bibliography{full}

%%%%%%%%%%%%%%%%%%%%%%%%%%%%%%%%%%%%%%%%%%%%
%%%%%%%%%%%%%%%%%%%%%%%%%%%%%%%%%%%%%%%%%%%%
\appendix

% %%%%%%%%%%%%%%%%%%%%%%%%%%%%%%%%%%%%%%%%%%%%%%%%%
\section{Adiabatic change in energy from a stellar flyby}
\label{sec:appendix}

Here we present the conclusions of \cite{Roy2003} and \cite{Heggie2006} for the change in energy of a binary star system due to the flyby of a third star.
The authors assume a hard binary, or that the binding energy of the binary far exceeds the kinetic energy of the passing star.
Although this traditional interpretation of a hard binary assumption is not consistent with planetary systems, the following analytical description for the change in energy of a star-planet system from a stellar flyby still holds for flybys in the adiabatic regime \citep[see the discussion in Section 2.1 by][]{Spurzem2009}.
The change in energy for a star-planet system due to an adiabatic flyby is given by \eq{eq:de}.
The functions $f_1(e_\star)$ and $f_2(e_\star)$ in that equation are
\begin{align}
    f_1(e_\star) &= \frac{(e_\star + 1)^{3/4}}{2^{3/4}\;e_\star^2}\,\mathrm{and}\\
    f_2(e_\star) &= \frac{3}{2\sqrt{2}}\frac{\sqrt{e_\star^2-1} - \arccos(1/e_\star)}{(e_\star-1)^{3/2}}\,.
\end{align}
For convenience, let us define the following functions of the planet's eccentricity $e$,
\begin{align}
    e_1 &= J_{-1}(e) - 2e J_0(e) + 2e J_2(e) - J_3(e) \\
    e_2 &= J_{-1}(e) - J_3(e) \\
  %  e_3 &= e J_{-1}(e) - 2 J_0(e) + 2 J_2(e) - e J_3(e) \\
    e_4 &= J_{-1}(e) - e J_0(e) - e J_2(e) + J_3(e)
\end{align}
where $J_n$ is the Bessel function of the first kind of order $n$.
With that notation,
\begin{align}
\begin{split}
    F(e, &i_\star, \omega_\star, \Omega_\star, n t_0) =\\
    &= e_1 \{ \sin(2\omega_\star + n t_0)\left[\cos(2i_\star) - 1\right] - \\
    &- \sin(2\omega_\star + n t_0)\cos(2i_\star)\cos(2\Omega_\star) - \\
    &- 3\sin(2\omega_\star + n t_0)\cos(2\Omega_\star) - \\
    &- 4\sin(2\Omega_\star)\cos(2\omega_\star + n t_0)\cos(i_\star) \} + \\
    &+ e_2 (1-e^2)\{ \sin(2\omega_\star + n t_0)\left[1 - \cos(2i_\star)\right] - \\
    &- \sin(2\omega_\star + n t_0)\cos(2i_\star)\cos(2\Omega_\star) - \\
    &- 3\sin(2\omega_\star + n t_0)\cos(2\Omega_\star) - \\
    &- 4\cos(2\omega_\star + n t_0)\sin(2\Omega_\star)\cos(i_\star)\} + \\
    &+ e_4 \sqrt{1-e^2}\{ - 2\cos(2i_\star)\cos(2\omega_\star + n t_0)\sin(2\Omega_\star) - \\
    &- 6\cos(2\omega_\star + n t_0)\sin(2\Omega_\star) - \\
    &- 8\cos(2\Omega_\star)\sin(2\omega_\star + n t_0)\cos(i_\star)\}\,,
\end{split}
\end{align}
where $n = \sqrt{G(m_c + m)/a^3}$ is the mean motion of the planet and $t_0$ is the time of pericentric passage of the planet where $t=0$ is when the flyby star passes perihelion.
The angles $\omega_\star, \Omega_\star,$ and $i_\star$ are defined for the flyby star with respect to the orientation of the star-planet system and are laid out in detail by \cite{Roy2003} in equation~(18).
Using the fact that the planet's semi-minor axis $b$ is related to its semi-major axis $a$ by $b = a\sqrt{1-e^2}$, we recover equation~(11) from \cite{Heggie2006}.

%%%%%%%%%%%%%%%%%%%%%%%%%%%%%%%%%%%%%%%%%%%%%%%%
\section{Angular momentum deficit}
\label{sec:amd}
The angular momentum deficit (AMD) of a multi-planet system is a measure of how eccentric and inclined the orbits of the planets are compared to a hypothetical system with the same mass and semi-major axis arrangement, but with coplanar circular orbits.
For an N-planet system in heliocentric coordinates, the AMD is defined as \citep{Laskar2000}
\begin{equation}
    C = \sum_{j=1}^{N} m_j\sqrt{G m_c a_j} \left(1 - \cos{i_j}\sqrt{1 - e_j^2}\right)
\end{equation}
where the planetary inclinations $i_j$ are defined with respect to the plane of the total angular momentum.
The measure of AMD can be used to determine stability, where a system is termed AMD-stable if the amount of AMD is insufficient to permit planet-planet collisions \citep{LaskarPetit2017}.

\begin{figure}
    \centering
    \resizebox{0.99\columnwidth}{!}{\includegraphics[trim=1cm 1cm 1cm 1cm]{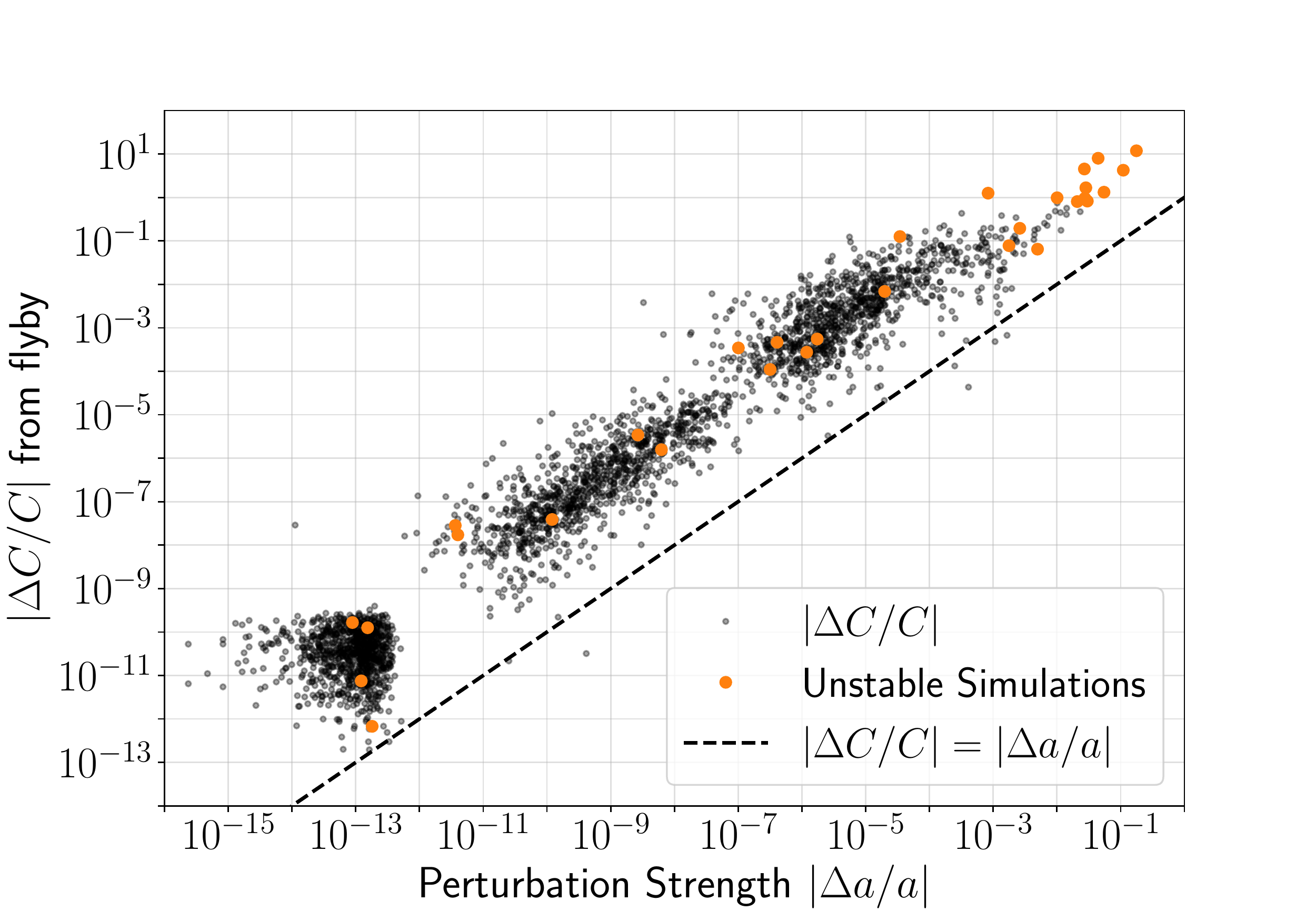}}
    \caption{Relative changes to the angular momentum deficit (AMD) of the entire Solar System (Sun and $8$ planets) $\Delta C/C$ as a function of the perturbation strength $\Delta a/a$, measured in Neptune's orbit all due to stellar flybys.
    \label{fig:amd}
    }
\end{figure}

In \fig{fig:amd} we plot the relative difference to the AMD $\Delta C/C$ versus the relative difference to Neptune's semi-major axis $\Delta a/a$, similar to \fig{fig:dg-vs-da}.
Although the AMD is slightly more sensitive at very small perturbation strengths ($\Delta a/a \lesssim 10^{-11}$), we find no difference in using the AMD to predict the long-term stability of the Solar System compared to using the changes in the semi-major axis of Neptune.
This might not be surprising, given the Solar System in its current state is already AMD-unstable \citep{Tamayo.et.al.2020}. 
The criterion for AMD-stability only indicates a worst-case scenario, whereas the probability of the Solar System being long-term stable is ultimately determined by secular dynamics.
This is why in the main paper, we decided to measure changes to semi-major axes and then relate them to changes in secular frequencies.

%%%%%%%%%%%%%%%%%%%%%%%%%%%%%%%%%%%%%%%%%%%%%%%%
\section{Details of numerical results}
\label{sec:details}

\begin{figure}
    \centering
    \resizebox{0.99\columnwidth}{!}{\includegraphics[trim=0.75cm 0.75cm 0.75cm 0.75cm]{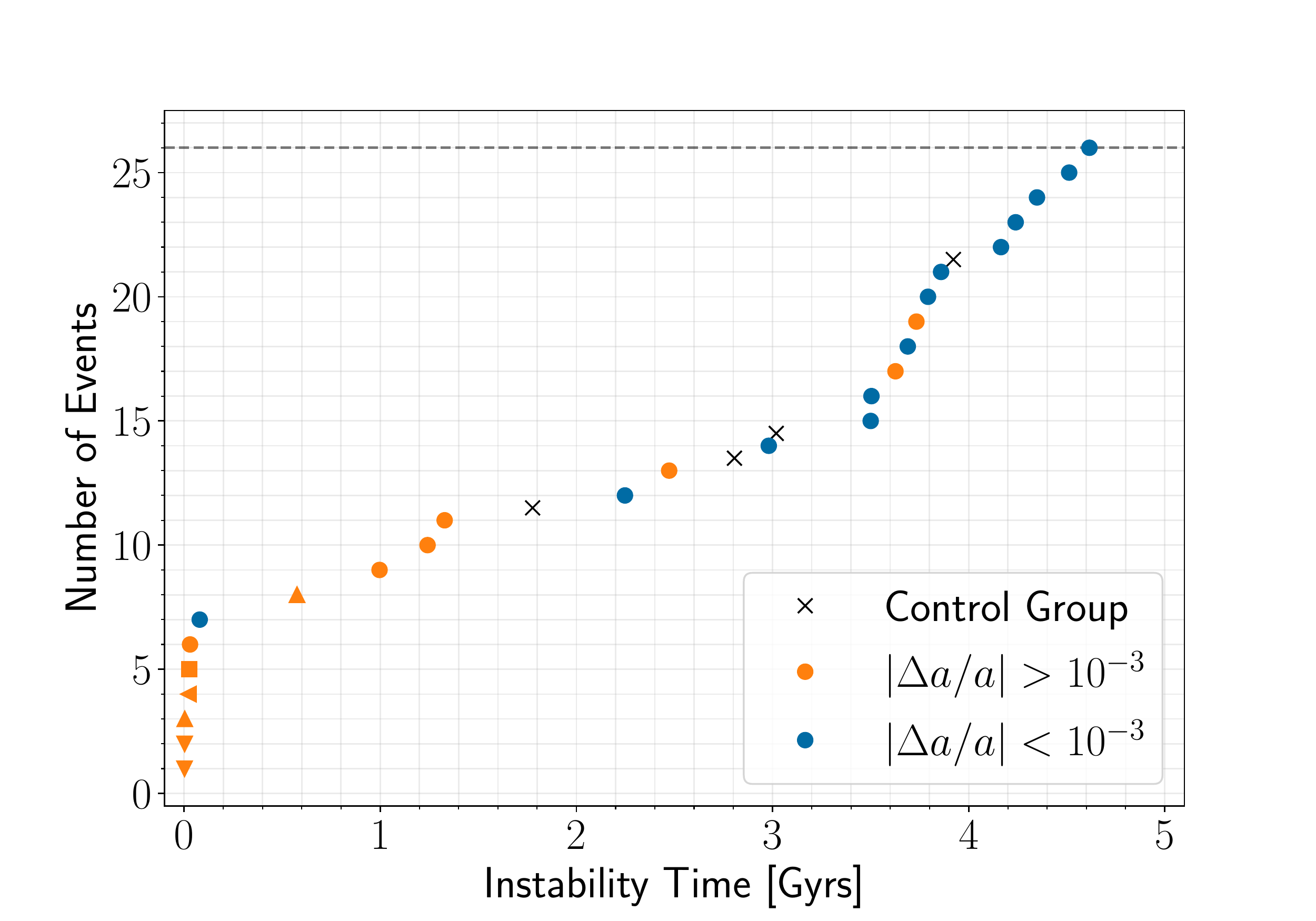}}
    \caption{All events that triggered the simulations from Section~\ref{sec:stability} to end before 5 Gyrs. 
    The color indicates if the initial perturbation is smaller or larger than $\tilde{a}_c = 10^{-3}$.
    The 24 events that were collisions between Mercury and Venus are indicated by circles ($\bigcirc$) for the experimental group and crosses for the control group ($\cross$).
    The square ($\square$) indicates a collision between Earth and Mars.
    The triangles indicate when the instability event was due to an escape rather than a collision.
    The triangle orientations indicate which planet escaped: Neptune ($\bigtriangleup$), Uranus ($\bigtriangledown$), and Mercury ($\triangleleft$).
    \label{fig:stability}
    }
\end{figure}

\fig{fig:stability} shows the cumulative number of simulations that go unstable with respect to the time that the instability event occurs.
There were 2,880 simulations in total and 30 instability events.
For all of the 24 events that were collisions between Mercury and Venus, there is a consistent correlation between a $g_1-g_5$ Mercury-Jupiter resonance and an increase in the eccentricity of Mercury preceding its collision with Venus.
All 4 of the 960 simulations in the control group and 20 of the 1,920 simulations in the experimental group ended with a Mercury-Venus collision.
The timing of these Mercury-Venus collisions are spread throughout the remaining lifetime of the Solar System, but there is a concentration of them around 3.5--4.5 Gyrs.

There was one simulation that ended with a collision between Earth and Mars and there is correlational evidence suggesting that the collision was preceded by an Earth-Mars secular resonance that pumped the eccentricities of both planets leading to a collision.
The remaining five events were escapes --- Uranus twice, Neptune twice, and Mercury once.
These early escapes of Uranus and Neptune are consistent with previous studies involving strong perturbations from stellar flybys \citep{Malmberg2011, Stock2020}.
The Uranus escapes were the result of extremely large perturbations from the flyby star which dramatically increased the eccentricity of Uranus and led to a close encounter with Saturn.
The Neptune escapes followed from a close encounter with Uranus and correlation evidence suggests that the close encounter was enabled by a series of eccentricity pumping from secular resonances with Uranus.
The path to Mercury's escape was complicated and is very difficult to disentangle.
We find correlations between various resonances on and off in different ways with many of the other planets coinciding with increases in Mercury's eccentricity and inclination.
All of this precedes Mercury experiencing a close encounter with the Sun and subsequent ejection from the system. 
Further investigation should be done into the correlations between secular resonances and eccentricity pumping within these simulations as the causal link to instability.

\end{document}